\documentclass[prl,preprint,superscriptaddress,showpacs,nobalancelastpage,nofootinbib]{revtex4}
\usepackage{graphicx,amsmath,amsbsy,amssymb}
\usepackage{color}

\newcommand{\ket}[1]{|{#1}\rangle} \newcommand{\bra}[1]{\langle{#1}|}

\begin{document}
  
  \title{Benchmarking  quantum control methods on a 12-qubit system}

\author{ C. Negrevergne} \affiliation{Institute for Quantum
  Computing, University of Waterloo, Waterloo, ON, N2L 3G1, Canada.}

\author{T.S. Mahesh} \affiliation{Department of Nuclear Engineering,
  MIT, Cambridge, Massachusetts 02139, USA} 

\author{C. A. Ryan} \affiliation{Institute for Quantum Computing,
  University of Waterloo, Waterloo, ON, N2L 3G1, Canada.}

\author{M. Ditty} \affiliation{Institute for Quantum Computing,
  University of Waterloo, Waterloo, ON, N2L 3G1, Canada.}

\author{F. Cyr-Racine} \affiliation{Institute for Quantum Computing,
  University of Waterloo, Waterloo, ON, N2L 3G1, Canada.}

\author{ W. Power} \affiliation{Institute for Quantum Computing,
  University of Waterloo, Waterloo, ON, N2L 3G1, Canada.}

\author{N. Boulant} \affiliation{Department of Nuclear Engineering,
  MIT, Cambridge, Massachusetts 02139, USA} 

\author{T. Havel}
\affiliation{Department of Nuclear Engineering, MIT, Cambridge,
  Massachusetts 02139, USA}

\author{D.G. Cory} \affiliation{Department of
  Nuclear Engineering, MIT, Cambridge, Massachusetts 02139, USA}

\author{R. Laflamme}
\affiliation{Institute for Quantum Computing, University of Waterloo, Waterloo, ON, N2L 3G1, Canada.}
\affiliation{Perimeter Institute for Theoretical Physics, Waterloo, ON, N2J 2W9, Canada}

\date{\today}


  \begin{abstract}
      In this letter, we present an experimental benchmark of
   operational control methods in quantum information processors
   extended up to 12 qubits. We implement universal control of this
   large Hilbert space using two complementary approaches and discuss
   their accuracy and scalability. Despite decoherence, we were able
   to reach a 12-coherence state (or 12-qubits pseudo-pure cat state),
   and decode it into an 11 qubit plus one qutrit labeled observable
   pseudo-pure state using liquid state nuclear magnetic resonance
   quantum information processors.

\end{abstract}

\maketitle

\section{Introduction}
Quantum mechanics promises information processors that are more
efficient than any known classical devices.  However, to bring this
potential to reality we must learn how to control large quantum
systems in a scalable fashion. Scalability has at least two
components: the complexity of the methods used to obtain coherent
control must grow only polynomially with the number of qubits
involved, and the errors occurring during the implementation of the
control sequence must be small enough to be correctable.  These errors
can be split in two classes: first the operational errors due to
imperfections in the control procedure and second, intrinsic errors
due to decoherence and relaxation processes.  Benchmarking small
Quantum Information Processor (QIP)
prototypes\cite{KLMT00a,FSB98,RHR04,BCS04} is therefore crucial to
characterizing the errors in a physical system and developing general
quantum control methods. In a physical system well suited for
implementing a QIP, once we have reached an optimal operational
control, one we will need to take care of intrinsic errors using
quantum error correction procedures \cite{DS96a}.

 Because they have the ability to run non trivial quantum algorithms,
liquid state Nuclear Magnetic Resonance (NMR) based
QIPs\cite{GC97a,CFH97a} can be used as benchmark systems
\cite{KLMT00a,VSB01,KLMN01a}.  In the present work we are interested
in optimizing operational control strategies in terms of accuracy and
amount of required classical resources.  To do so we have chosen to
extend the benchmarking algorithm previously used on a 7-qubit liquid
state NMR register\cite{KLMT00a}, to 11-qubit-plus-one-qutrit (see
Fig.1).  This algorithm consists of preparing mixtures of generalized
GHZ states of the form: $\rho_{GHZ} = \textbf{I}^{\otimes n} +
\textbf{X}^{\otimes n}$ ($\textbf{I}$ is the identity matrix,
$\textbf{X}$ is the $\boldsymbol{\sigma_x}$ Pauli matrix and $n$ is
the number of qubits involved in the GHZ states).  This state
preparation is very similar to the generation of stabilizer operators
\cite{Got97a}, which are building-blocks for quantum error correction
codes. Furthermore, this algorithm takes the state of the quantum
system to the most fragile reaches of the Hilbert space we are
operating in, and therefore clearly demonstrates coherent control.
Previous work has demonstrated a 12-spin pseudo-cat state \cite{LK05a}
and multiple quantum coherences of much higher order\cite{WWP80a}.
However, these exploited symmetries in the systems which limited them
to a much smaller symmetric subspace of the full Hilbert space.  In
the present Letter we benchmark \emph{universal} control methods that
allow us to access the full Hilbert space of our system.\\

\section{Coherent universal control schemes}
  In liquid state NMR QIP, universal control is achieved through the
application of a coordinated sequence of radio-frequency (RF) pulses
and periods of free evolution. The resulting one and two-qubit gates
allow us, in principle, to implement any unitary transformation
\cite{BBC+95a}. The challenge is to efficiently design such pulse
sequences to be as short as possible and robust against experimental
imperfections in order to minimize systematic error and decoherence
\cite{KBG01,JJ03}. In a three-qubit experiment \cite{CPM98} it is
possible to write the pulse sequences by hand and compensate for
experimental errors with a few optimization parameters. Moving to
larger registers \cite{KLMT00a,VSB01} requires more complex control
schemes that necessitate systematic numerical optimization in the
design of the pulse sequence. Going to 12-qubits represents a
substantial step forward in the number of quantum degrees of freedom
that are controlled.

  We will approach coherent control over this large Hilbert space
system from two complementary points of view.  First, to demonstrate
that control methods of high precision are available and
experimentally realizable, we build a detailed model of the
experimental QIP and for each desired unitary operation, search for an
optimal control sequence based on strongly modulating pulses
\cite{FMP02}.  Applied over the entire Hilbert space, this approach is
not scalable. The amount of classical resources required to search for
control sequences grows as the size of the Hilbert space -
i.e. exponentially with the number of qubits.  However, this approach
returns control sequences of high fidelity and with small, known
errors, provided our system model is accurate. Because of the
exponential computing cost of determining a suitable pulse, when
dealing with a large Hilbert space we have to lower the dimension of
the space over which we search. This can be achieved by searching for
pulses only on a sub-system of the spins of the register (in the
present case the carbon nuclear spins) and check that it leads to
sufficiently high fidelity control by simulating the effect of rest of
the spins as described in \cite{WHE04}.

 A second approach to control such a large Hilbert space is to make a
series of well-constructed simplifications to the model, in order to
permit control sequences to be developed with a complexity that grows
only polynomially with the number of qubits. We therefore based the
design of our pulse sequences on the method developed in
\cite{KLMT00a}.  Indeed, by using only simple pulses (broad-band
rectangular hard pulses and selective soft Gaussian shaped pulses) and
performing a series of simulations on pairs of spins with significant
couplings for each pulse, it is possible to efficiently determine
first order deviations from the ideal pulse. For each of the pulses,
these control errors can be represented as phase shifts and spin-spin
coupling effects occurring before and after an ideal pulse. One can
then modify the phase of each pulse to correct for the phase shifts.
Assuming that long range couplings between the spins vanish, the
timing between pulses can be efficiently numerically optimized in
order to absorb the coupling effects into the refocusing
scheme\cite{BJLK05}. This design does not take higher-order coupling
and off-resonant effects into account and leave some small couplings
un-refocused to minimize the pulse sequence length. These
approximations lead to errors in the control. A crucial point of this
experimental work was to verify that these approximations hold for
larger Hilbert spaces i.e. that we could find a suitable refocusing
scheme that, once optimized, still provides reliable control.\\

\section{Experimental results and discussions}
In liquid state NMR, the thermal equilibrium state is almost
completely mixed.  Therefore, instead of preparing $\rho_{GHZ}$ we
actually prepare the following state:
  \begin{equation}
    \rho_{cs} \simeq \frac{\textbf{I}^{\otimes N}}{2^{N}} + \epsilon
    \textbf{X}^{\otimes n}\textbf{I}^{\otimes (N-n)},
  \end{equation} 
$N=14$ is the total number of spins-$1/2$ in the register, and $n$ is
the number of qubits involved in the GHZ state. The factor $\epsilon
\simeq 10^{-5} $ is related to the thermal polarization of the system.
The second term of $\rho_{cs}$ , called the deviation density matrix,
contains the n-coherence term
$\ket{00...0}\bra{11...1}+\ket{11...1}\bra{00...0}$ corresponding to a
n-qubit cat-state $\ket{00...0}+\ket{11...1}$, as well as lower
coherence terms corresponding to the other n-qubit GHZ states. This
state preparation (called the encoding of the pseudo-cat state) is
done by propagating the polarization of the two equivalent protons
through the chain of nuclei by a sequence of one and two-qubit quantum
gates (see Fig. 1). In NMR only single coherence terms are
observable\cite{EBW94a}. Therefore, to see the signature of the GHZ
state we need transform the n-coherence term into a n-qubit labeled
pseudo-pure state of the form $\textbf{X00....0}$ (where
$\textbf{0}=(\textbf{I}+\boldsymbol{Z})/2$). This step of the
algorithm is called the decoding.  To average away the other lower
coherence order terms present in the $\textbf{X}^{\otimes n}$
operator, we used two types of coherence filters: gradient and phase
cycling techniques. Proof that we have actually created the
pseudo-pure and accompanying pseudo-cat state by determining the final
state through tomography would require an impractically large number
of experiments ($\sim 4^{12}$). Nevertheless, since the averaging
procedure filters out signal coming from every term but the desired
one (i.e. the highest coherence order term), a single observation of
the ``read out'' nucleus in the resulting NMR spectrum (see Fig. 1)
indicates that we have indeed reached the desired coherence.

We applied both methods to design two series of pulse sequences that
implement the encoding-decoding procedure, with n going from 1 to 12,
on a liquid state NMR QIP, based on uniformly $^{13}C;^{15}N$ labeled
l-histidine (See Fig. 2).  Two different samples were used. The one
used for designing strongly modulating pulses was made of 16.7 mg of
histidine, 15.9 mg of deuterated phosphoric acid in 1ml of deuterated
water.  To design simplified pulse sequences we prepared a second
sample by dissolving 35.3 mg of histidine, 12.5 mg of
glycine-$2-^{13}C;^{15}N$ and 3.4 mg of deuterated phosphoric acid in
1 ml of deuterated water. The labeled glycine molecule has a simple
spectrum which allowed us to perform accurate calibrations of the
selective pulses on isolated NMR peaks \emph{in situ}. The experiments
based on the strongly modulating pulses and the simplified design were
respectively performed on Avance-600 and Avance-700 Bruker
spectrometers at MIT and IQC.

We based the design and the interpretation of the experiments on a
model of the system and the apparatus\cite{WHE04} which includes the
following attributes: (1) The Hamiltonian of the system in the static
magnetic field of the spectrometer. The chemical shifts as well as the
scalar-coupling strengths and relatives signs were experimentally
determined by fits of reference spectra and small targeted multiple
coherence experiments. (2) Knowledge of $T_2$ and $T_2^*$
\cite{EBW94a} relaxation times of the system. (3) RF field
inhomogeneities were mapped and used in the design the strongly
modulating pulses \cite{PBE03}.

 This type of experiment comes with a predicted exponential decay of
signal as we increase the number of correlated qubits. We also expect
high decoherence rates \cite{KS04,LK05b} and therefore a strong signal
attenuation, as it is reasonable to assume that the relaxation rate
for each spin included in the multiple quantum coherence add. To
evaluate the quality of the control we could reach, the relevant
quantity to measure is the amount of signal obtained experimentally
with respect to the expected one assuming perfect control. Fig. 3
shows how much signal we were able to retain after decoding the
highest coherence order cat state into a pseudo-pure state for each
experiment. We could reach a 12-coherence state using strongly
modulating pulses and a 10-coherence state with selective
pulses. Indeed, the sequences obtained through the simplified design
were slightly longer, leading to more decoherence. Moreover, the
transverse relaxation times were not the same in both sets
of experiments. To distinguish between operational-errors and
relaxation loss, both decay times ($T_2^*$ and $T_2$) were used to
estimate the signal loss due to transversal relaxation during the
pulse sequences (See Fig.4). It showed that decoherence is the main
source of signal loss and therefore indicates that we have good
operational control.\\

\section{Conclusion}
  In summary, we have reported an algorithmic benchmark performed on
  the largest quantum information processor to date. Despite the
  decoherence during the computation, we have been able to demonstrate
  universal control on up to 11 qubits and one qutrit.  This work
  shows that liquid state NMR allows us to develop operational control
  methods that can be used to control a large number of quantum
  degrees of freedom. These methods provide a systematic and efficient
  way of programming liquid state NMR QIPs. However, the approaches
  and control techniques behind these methods could also be used to
  design control sequences in more scalable implementations where the
  intrinsic errors are smaller.

This work was supported by ARDA, ARO, LPS, NSERC and by the
Cambridge-MIT Institute.

\newpage


\begin{figure}[htbp]
 \begin{center}
 \includegraphics[width=8.2cm,height=4.5cm]{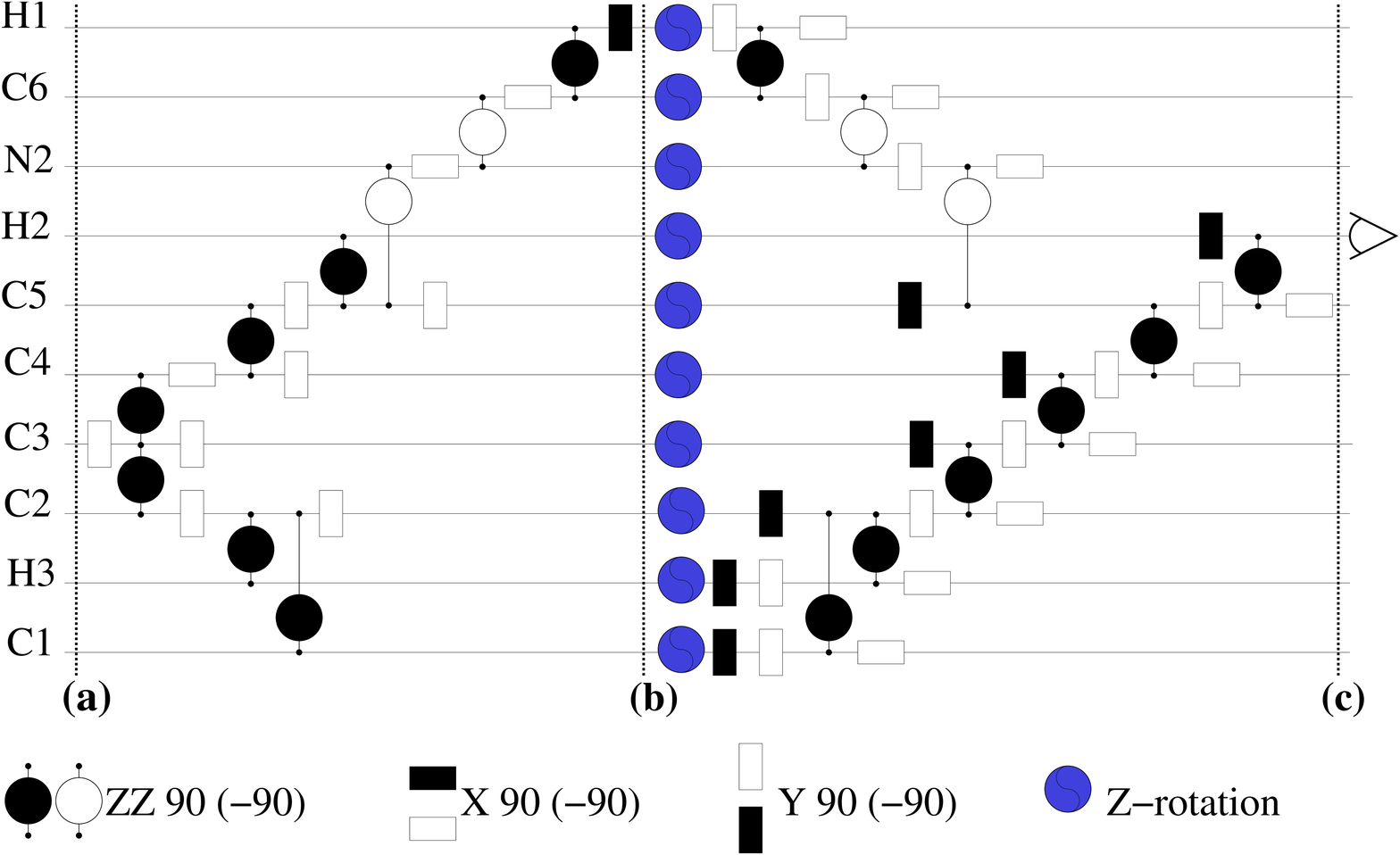}
 \end{center}
 \caption{Sequence of gates for the 10-qubit pseudo-cat state
 preparation followed by its decoding into a 10-qubit pseudo-pure
 state. The initial preparation of the qutrit into a pseudo-pure
 state, as well as the refocusing gates, are not shown. Proper cycling
 of the Z-rotations and the phase of observation act as a coherence
 filter. The qubits names correspond to the histidine molecule nuclei
 (see Fig.2) a) After the qutrit pseudo-pure preparation, the state of
 the register is $\textbf{0}^{H_{4/5}}
 \textbf{I}^{H_1}\textbf{I}^{C_6}\textbf{I}^{N_2}\textbf{I}^{H_2}\textbf{I}^{C_5}\textbf{I}^{C_4}\textbf{Z}^{C_3}\textbf{I}^{C_2}\textbf{I}^{H_3}\textbf{I}^{C_1}$
 . At the end of the encoding in b) it is $\textbf{0XXXXXXXXXX}$ and
 after filtering, the decoded state in c) is $\textbf{0000X000000}$. }
 \end{figure}

\begin{figure}[htbp]
\includegraphics[width=8.4cm,height=4cm]{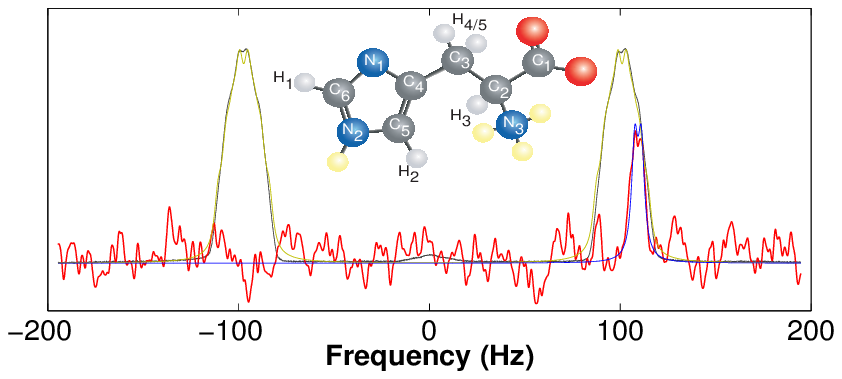}
\caption{This l-histidine molecule has 14 spin-1/2 nuclei: five
$^{1}H$, six $^{13}C$ and three $^{15}N$. See EPAPS Document No.1 for
a more complete description of the molecule.The two protons $H_{4}$
and $H_{5}$, are chemically equivalent and indistinguishable. As such,
they can be seen as an composite particle with a spin-1 and a spin-0
part. We considered only the spin-1 sub-space (qutrit) since the
spin-0 does not interact with the other spin-1/2. This molecule is
therefore a 12-qubits plus one qutrit quantum register. However, the
$N_3$ nuclear spin has a particularly weak coupling with the rest of
the molecule; thus we did not use it. On this plot we have shown a
reference spectrum of $H_2$ (gray plot) and of the pseudo-pure state
obtained after decoding a 10-qubit cat-state onto $H_2$ (red
line). They are arbitrarily scaled for clarity. The reference spectrum
was obtained with 2 scans and the pseudo-pure, with 4000 scans, in
order to improve the signal to noise ratio. We also show simulated
spectra of the expected reference (yellow plot) and pseudo-pure state
(blue line) for which the amplitude is matched to the experimental
data to evaluate the signal loss.  }
 \end{figure}

\begin{figure*}[htbp]
\begin{center}
\includegraphics[width=16cm]{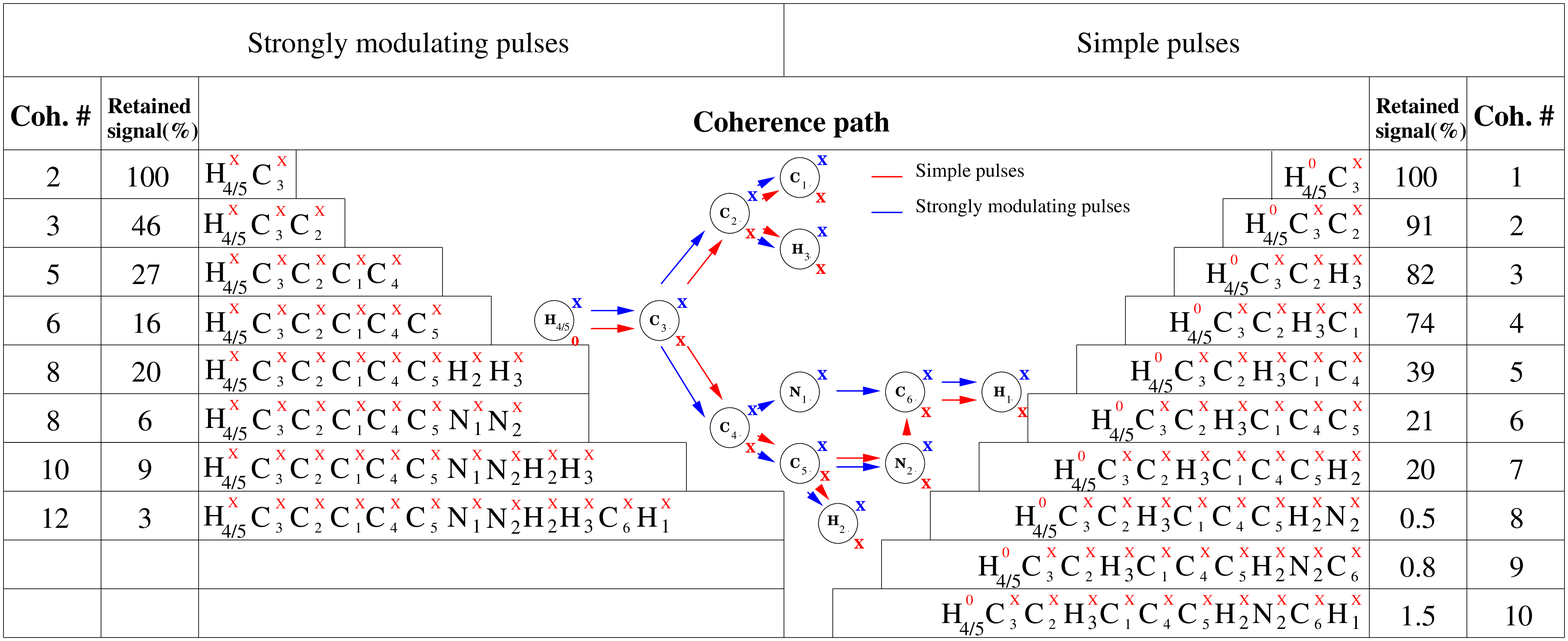}
\end{center}
\caption{Description of each of the multiple-coherence
experiments. The picture shows how the polarization of the $H_{4/5}$
is propagated through the nuclei chain to create the cat state. In the
series of experiments using simple pulses we first prepared the qutrit
made of the two equivalent protons into a pseudo-pure state
$\textbf{0}= (\textbf{I}+\boldsymbol{Z})/2$ and left it as such for
the rest of the experiments. When using strongly modulating pulses, we
included the qutrit into the multiple coherence state.  The two first
and two last columns show the coherence number we reached and how much
signal we where able to retain after decoding the cat-state into a
observable pseudo-pure state. Results are shows in \% of the expected
signal assuming perfect control, normalized to the first experiment. }
\label{results}
\end{figure*}

\begin{figure}[htbp]
\begin{center}
\rotatebox{270}{\includegraphics[width=4.5cm,height=8.2cm]{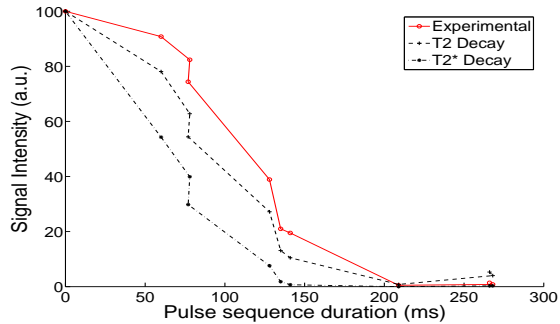}}
\end{center}
\caption{ Expected decay of the pseudo-pure state signal due to
transversal relaxation for the series of experiments done with simple
pulses. Each point correspond to a different coherence experiment. The
length of the pulse sequence increases with the coherence order.  Most
experimental points are above the estimates given by $T_2^*$ and even
$T_2$. Indeed, to predict the transversal relaxation during the pulse
sequence, we only used a very simple model of decay for multiple
coherences that gives an upper-bound of the signal loss. Nevertheless,
the experimental curve and predicted ones show the same decay
pattern. Thus it is reasonable to say that most of the signal loss
comes from decoherence and therefore that we have a good operational
control over the system. For the coherences 8, 9 and 10 (three last
points), the experimental curve goes below the $T_2$ curve. It
reflects a loss of accuracy in our operational control. Indeed, for
these experiments, we are controlling the nitrogen nuclei through very
small couplings. We are therefore using long two-qubits gates that are
sensitive to any small inaccuracy in the values of the Hamiltonian
parameters.  }
\end{figure}

\end{document}